\documentclass[prbr,aps,twocolumn,groupedaddress,showpacs,letterpaper]{revtex4}

\usepackage{bm}
\usepackage{graphicx}
\usepackage{color}
\usepackage{ulem}
\usepackage[top=20truemm,bottom=8truemm,left=18truemm,right=18truemm]{geometry}

\begin{document}

\title{Observation of Dirac-like energy band and unusual spectral line shape in quasi-one-dimensional superconductor Tl$_2$Mo$_6$Se$_6$}

\author{Kosuke Nakayama,$^1$ Zhiwei Wang,$^2$ Chi Xuan Trang,$^1$ Seigo Souma,$^{3,4}$ Emile D. L. Rienks,$^{5,6}$ Takashi Takahashi,$^{1,3,4}$ Yoichi Ando,$^2$ and Takafumi Sato$^{1,3}$}

\affiliation{$^1$Department of Physics, Tohoku University, Sendai 980-8578, Japan\\
$^2$Institute of Physics II, University of Cologne, D-50937 K\"oln, Germany\\
$^3$Center for Spintronics Research Network, Tohoku University, Sendai 980-8577, Japan\\
$^4$WPI Research Center, Advanced Institute for Materials Research, Tohoku University, Sendai 980-8577, Japan\\
$^5$Institute for Solid State Physics, Leibniz Institute for Solid State and Materials Research Dresden, D-01171 Dresden, Germany\\
$^6$Institute of Solid State Physics, Dresden University of Technology, D-01062 Dresden, Germany
}

\date{\today}

\begin{abstract}
We have performed high-resolution angle-resolved photoemission spectroscopy of the quasi-one-dimensional (1D) topological superconductor candidate Tl$_2$Mo$_6$Se$_6$ consisting of weakly-coupled Mo$_3$Se$_3$ chains. We found a quasi-1D Fermi surface arising from a Dirac-like energy band, which is associated with the nonsymmorphic screw symmetry of the chains and predicted to trigger topological superconductivity. We observed a significant spectral-weight reduction over a wide energy range, together with a tiny Fermi-edge structure which exhibits a signature of a superconducting-gap opening below the superconducting-transition temperature. The observed quasi-1D Dirac-like band and its very small density of states point to an unconventional nature of superconductivity in Tl$_2$Mo$_6$Se$_6$.
\end{abstract}


\maketitle
The interplay between low dimensionality and superconductivity is one of the central issues in condensed-matter physics. The reduction of dimensionality triggers the rearrangement of lattice, charge, spin, and orbital degrees of freedom, leading to the competition or cooperation among various ordered phases as well as phase fluctuations. Since these rich properties and their interplay often give rise to unconventional superconductivity beyond the Bardeen-Cooper-Schrieffer (BCS) theory, low-dimensional electron systems are a fertile playground for the research of exotic superconductivity. Well-known examples of low-dimensional superconductors include quasi-two-dimensional (2D) copper oxides and iron pnictides/chalcogenides which show superconductivity at unexpectedly high temperatures \cite{Cuprate, Pnictide}. Interfaces and atomically thin films \cite{Interface, Film}, regarded as the ultimate of 2D materials, are now intensively investigated for the search of novel superconductivity that cannot be found in quasi-2D materials. On the other hand, superconductivity in one-dimensional (1D) or quasi-1D materials is relatively scarce, because of the difficulty in realizing a superconducting state in such systems owing to the Peierls instability as well as strong fluctuations.

\begin{figure}
\includegraphics[width=3.4in]{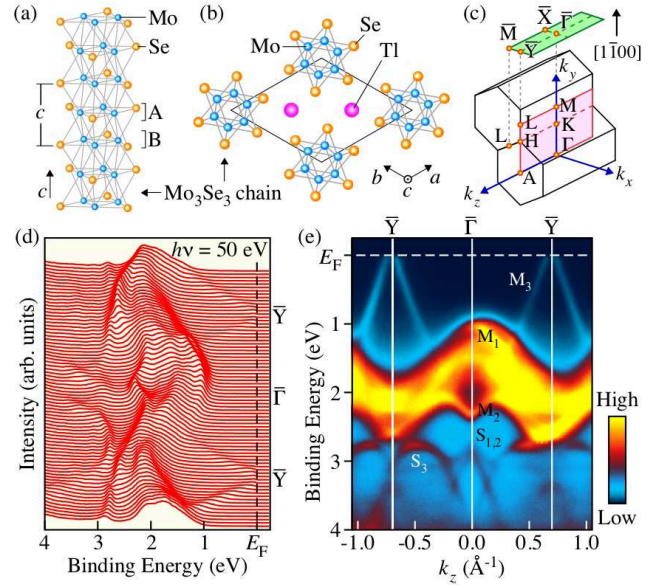}
\vspace{-0.5cm}
\caption{(a) 3D view of a Mo$_3$Se$_3$ chain running along the $c$-axis. (b) Top view of the crystal structure. (c) Bulk hexagonal BZ (black) and corresponding surface BZ projected onto the (1$\bar 1$00) plane (green rectangle). (d) and (e) ARPES spectra and corresponding intensity plot in the VB region, respectively, measured at $T$ = 40 K.}
\end{figure}

\begin{figure*}
\includegraphics[width=6.8in]{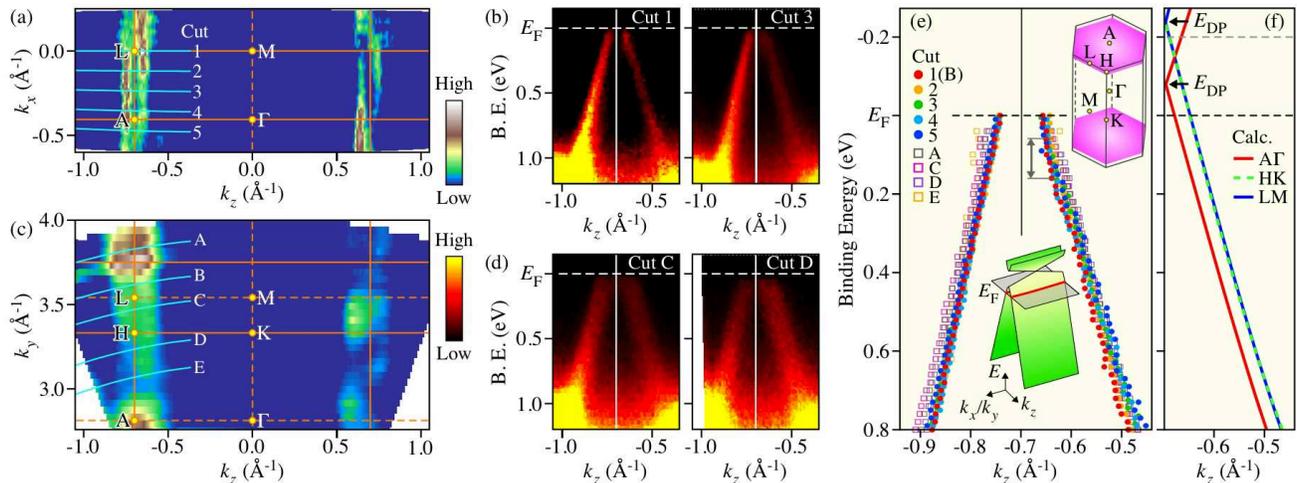}
\vspace{-0.5cm}
\caption{(a) ARPES intensity at $E_{\rm F}$ plotted as a function of $k_x$ and $k_z$ measured with $h$$\nu$ = 50 eV. (b) ARPES intensity plotted as a function of $k_z$ and binding energy measured along cuts 1 and 3 indicated by blue lines in (a). (c) Same as (a) but plotted as a function of $k_y$ and $k_z$ obtained by sweeping photon energy $h$$\nu$. Measured cuts A-E with $h$$\nu$ = 56, 50, 46, 40, and 36 eV, respectively, are indicated by blue lines \cite{innerpotential}. (d) Same as (b) but measured along cuts C and D in (c). (e) Comparison of VB dispersions for cuts 1-5 and A-E in (a) and (c), respectively. The upper right inset shows a schematic FS expected from the experimental band dispersion. The lower inset shows a schematic Dirac-cone dispersion. (f) Dirac-like band dispersions obtained from band calculations \cite{TSC}. To take into account the finite hole doping in our sample, the energy position of $E_{\rm F}$ was set to be 200 meV below $E_{\rm F}$ in the band calculations (gray dashed line).}
\end{figure*}

Among known quasi-1D superconductors \cite{Organic,LiMo7O11,A2Cr3As3}, molybdenum selenide Tl$_2$Mo$_6$Se$_6$ \cite{TMS1, TMS2} is an ideal platform for studying quasi-1D superconductivity because of a high chemical stability, a relatively high superconducting transition temperature ($T_{\rm c}$ = 6.5 K), and a strong 1D anisotropy in the normal and superconducting states as seen in the electrical resistivity and upper critical field \cite{TMS1, Anisotropy1, Anisotropy2, Calc1}. A basic unit of the crystal structure in Tl$_2$Mo$_6$Se$_6$ is an infinitely long Mo$_3$Se$_3$ chain situated along the $c$ axis [Fig. 1(a)]. Each 1D chain is separated by Tl atoms which prevent direct chemical bonding between adjacent chains [Fig. 1(b)]. The resultant weak inter-chain interaction provides the crystal a quasi-1D character. The quasi-1D electronic properties of Tl$_2$Mo$_6$Se$_6$ provide a rare opportunity to search for the Tomonaga-Luttinger-liquid (TLL) state. Furthermore, in addition to these basic interests in the 1D properties, Tl$_2$Mo$_6$Se$_6$ is now attracting great attention as a topological material candidate \cite{DSM, Cubic, TSC}. In fact, first-principles band-structure calculations suggested that Tl$_2$Mo$_6$Se$_6$ is a Dirac semimetal characterized by a nonsymmorphic symmetry \cite{DSM, Cubic}, which originates from the presence of two sublattices (A and B) connected to each other by a two-fold screw operation [Fig. 1(a)]. In particular, one of the Dirac-cone energy bands represents a novel Dirac fermion termed a {\it cubic Dirac fermion} which shows a cubic (linear) dispersion perpendicular (parallel) to the chain \cite{Cubic}. Further, a very recent theory proposed that Tl$_2$Mo$_6$Se$_6$ may realize time-reversal-invariant topological superconductivity associated with the symmetry-protected Dirac cones even in the absence of spin-orbit coupling \cite{TSC}. To examine the origin of quasi-1D properties and theoretical predictions for Dirac-semimetal and topological-superconductor phases, the experimental determination of the electronic structure of Tl$_2$Mo$_6$Se$_6$ is urgently required.

In this Rapid Communication, we report high-resolution angle-resolved photoemission spectroscopy (ARPES) of Tl$_2$Mo$_6$Se$_6$ single crystal. By utilizing energy-tunable photons from synchrotron radiation, we obtained definitive evidence for the formation of a quasi-1D Fermi surface (FS) originating from the linearly-dispersive Dirac-like band. We also observed a significant suppression of the density of states (DOS) near the Fermi level ($E_{\rm F}$) which coexists with superconductivity. We discuss the implications of our observations in relation to the superconducting and topological properties.

High-quality Tl$_2$Mo$_6$Se$_6$ crystals with $T_{\rm c}$ = 6.3 K were grown from elemental Tl shot (purity 99.99\%), Mo powder (purity 99.999\%), and Se powder (purity 99.999\%) \cite{Wang,SM}. ARPES measurements were performed at UE112\_\,PGM-2b-1$^3$ beamline of BESSY, Helmholtz-Zentrum Berlin. The ARPES end station is equipped with an Omicron-Scienta R4000 electron analyzer with a Janis 1-K cryostat which enables access to the superconducting phase of Tl$_2$Mo$_6$Se$_6$. The energy and angular resolutions were set at 4-30 meV and 0.3$^{\circ}$, respectively. Samples were cleaved {\it in situ} along the (1$\bar 1$00) crystal plane in an ultrahigh vacuum of better than 1$\times$10$^{-10}$ Torr. Corresponding bulk and surface Brillouin zones (BZs) are shown in Fig. 1(c). Details of the sample geometry and the core-level spectrum are described in the Supplemental Material \cite{SM}.

To discuss the valence-band (VB) structure, we performed ARPES measurements along the $\bar \Gamma$$\bar {\rm Y}$ cut (parallel to the Mo$_3$Se$_3$ chain) of the surface BZ with $h$$\nu$ = 50 eV [Figs. 1(d) and 1(e)]. At a binding energy ($E_{\rm B}$) = 2.5-4.0 eV, one can find several dispersive bands such as two holelike bands (S$_{1,2}$) at the $\bar \Gamma$ point and another holelike band (S$_{3}$) with the top of the dispersion slightly away from the $\bar {\rm Y}$ point. According to the first-principles calculations, these bands have a dominant Se 4$p$ character with a finite admixture of Mo 4$d$ \cite{Calc1, Calc2}. More intense features are observed at $E_{\rm B}$ = 1.0-2.5 eV (M$_1$ and M$_2$), which mainly stem from the bonding states of Mo 4$d$ orbitals. The antibonding counterparts of the Mo 4$d$ states contribute to conduction bands above $E_{\rm F}$ \cite{Calc1, Calc2}. In contrast to the presence of multiple band dispersions at $E_{\rm B}$ = 1.0-4.0 eV, the electronic structure near $E_{\rm F}$ is rather simple. One can find a single highly dispersive band (M$_3$) which mainly arises from the bonding states of the Mo 4$d_{xz}$ orbital \cite{Calc1, Calc2}.

To see more clearly the electronic states near $E_{\rm F}$, we mapped out the band structure in the 3D bulk BZ. In Figs. 2(a) and 2(b), we show FS in the $k_x$-$k_z$ plane and representative band dispersions around the BZ corner, respectively. As seen in Fig. 2(b), the $d_{xz}$ band linearly disperses and crosses $E_{\rm F}$ in a momentum region slightly away from the $\bar {\rm Y}$$\bar {\rm M}$ line, consistent with the metallic property of Tl$_2$Mo$_6$Se$_6$. More importantly, the Fermi wave vector ($k_{\rm F}$) is stationary to the variation of $k_x$, as recognized from a nearly straight FS shape in Fig. 2(a) (note that there is a finite ambiguity for the position of $k_x$ = 0 because of the lack of periodic band dispersion as a function of $k_x$). Large anisotropy of the electronic structure is also observed in the $k_y$-$k_z$ plane [Figs. 2(c) and 2(d)]; namely, there is no large FS warping nor clear change in the band dispersion along $k_y$. To quantitatively discuss the anisotropy of the electronic structure, we compared in Fig. 2(e) the band dispersions extracted at various $k_x$ and $k_y$ cuts. Apparently, the linear band dispersion is robust against the variations of $k_x$ and $k_y$. Linear fittings to the band dispersions at various cuts give only small changes in $k_{\rm F}$ (0.65-0.67 ${\rm \AA}$$^{-1}$) and the Fermi velocity $v_{\rm F}$ (3.8-4.7 eV${\rm \AA}$). The magnitude of the band dispersion along the $k_x$ and $k_y$ axes is at most $\sim$100 meV, as estimated from the energy difference of band dispersions at different $k_x$ and $k_y$ cuts [see an arrow in Fig. 2(e)]. This value is ten times smaller than that along $k_z$ ($>$ 1 eV). These results unambiguously demonstrate the quasi-1D nature of the electronic structure in Tl$_2$Mo$_6$Se$_6$. Since other bands except for the quasi-1D band do not cross $E_{\rm F}$, the FS of Tl$_2$Mo$_6$Se$_6$ consists only of almost flat sheets near the BZ boundary, as illustrated in the upper right inset to Fig. 2(e). The carrier concentration estimated from the FS volume is 3.7 $\times$ 10$^{20}$ cm$^{-3}$, which shows a good agreement with that from the Hall coefficient ($\sim$3 $\times$ 10$^{20}$ cm$^{-3}$) \cite{Anisotropy2}.

The observed quasi-1D electronic structure shows a reasonable agreement with band structure calculations \cite{Calc1, DSM, Cubic, TSC, Calc2} when the chemical potential is located $\sim$200 meV below the calculated $E_{\rm F}$ (likely due to Tl deficiencies) [compare Figs. 2(e) and 2(f)]. In particular, $k_{\rm F}$ and $v_{\rm F}$ show a quantitatively good agreement between the experiment and calculations, indicating the bulk nature of the observed band structure. Moreover, our observation of a linearly dispersive feature is a signature of the non-trivial electronic structure in Tl$_2$Mo$_6$Se$_6$. According to the theoretical calculations \cite{Calc1, DSM, Cubic, TSC, Calc2}, the two-fold screw symmetry of the Mo$_3$Se$_3$ chain forces the degeneracy of the bonding and antibonding Mo 4$d_{xz}$ bands at the A and H points [the resulting Dirac points are marked by black arrows in Fig. 2(f)] and hence results in a Dirac-semimetal phase with anisotropic Dirac-cone-like dispersions near $E_{\rm F}$. By linearly extrapolating the experimental band dispersions (lower branch of the Dirac cone), we estimate the energy position of the Dirac point ($E_{\rm DP}$) to be 100-200 meV above $E_{\rm F}$. The observed quasi-1D electronic structure indicates a highly anisotropic nature of the Dirac-cone dispersion: as illustrated in the lower inset to Fig. 2(e), the Dirac cone would have a relatively large dispersion in the $k_z$ direction compared with the negligible  $k_x$/$k_y$ dispersion. This large anisotropy corresponds to the $k_{\rm F}$ value almost independent of $k_x$/$k_y$, creating sheet-like FSs (red lines in the lower inset).

\begin{figure}
\includegraphics[width=3.4in]{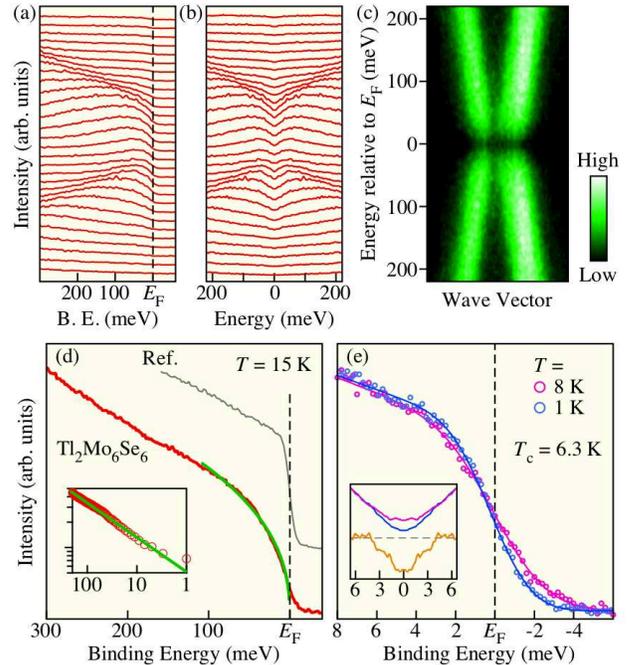}
\vspace{-0.5cm}
\caption{(a) and (b) ARPES spectra and corresponding symmetrized spectra, respectively, measured at $T$ = 30 K along cut 1 in Fig. 2(a). (c) Intensity plot of (b). (d) Simulated DOS at $T$ = 15 K (red) obtained by integrating the ARPES intensity between $k_z$ = -0.5 and -0.9 ${\rm \AA}^{-1}$ [comparable to the whole range of the horizontal axis in (c)]. Green curve shows a result of power law fitting to DOS. Fermi-Dirac cutoff of a referenced metal is plotted by gray curve. The inset shows a logarithmic plot of DOS (red) and a result of linear fitting (green). (e) Comparison of near-$E_{\rm F}$ DOS at $T$ = 1 K and 8 K recorded with our highest energy resolution $\Delta$$E$ = 4 meV with $h$$\nu$ = 18 eV. Solid curves show numerical fitting results. The inset shows symmetrized DOS at 1 K and 8 K (magenta and blue, respectively) and their relative change (yellow) obtained by dividing the symmetrized DOS at 1 K by that at 8 K (unity is indicated by gray dashed line).}
\end{figure}

A hallmark of a quasi-1D electron system appears in the spectral line shape. Figures 3(a)-3(c) display representative ARPES spectra measured around the $k_{\rm F}$ point at $T$ = 30 K, their symmetrized curves, and corresponding intensity plot, respectively. As recognized from Figs. 3(a) and 3(b), a quasiparticle peak is not well defined at the $k_{\rm F}$ point, indicating a marked suppression of spectral weight in the normal sate [also see Fig. 3(c)]. Correspondingly, DOS [red curve in Fig. 3(d)] starts to decrease at $\sim$50 meV and is continuously reduced with approaching $E_{\rm F}$ in a power-law-like manner ($\propto$$|E|^{0.42}$; see the green curve and the inset), in contrast to a steep Fermi edge in a 3D metal (gray curve). Such a suppression of DOS over a relatively wide energy range is characteristic of quasi-1D materials \cite{ARPES1, ARPES2, ARPES3, ARPES4, ARPES5}. The observed power-law-like suppression persists down to $T$ = 1 K (not shown) well below $T_{\rm c}$, suggesting the coexistence with superconductivity. It is remarked here that while the DOS is significantly suppressed, our high-resolution measurements in close vicinity of $E_{\rm F}$ reveal a very small but finite step edge at low temperatures [Fig. 3(e)]. This remaining tiny DOS at $E_{\rm F}$ must be responsible for superconductivity in Tl$_2$Mo$_6$Se$_6$. In fact, we found a finite leading-edge shift by $\sim$0.2 meV across $T_{\rm c}$ which suggests the superconducting-gap opening [Fig. 3(e)]. A signature of gap opening below $T_{\rm c}$ is more clearly seen by comparing the symmetrized spectra shown in the inset to Fig. 3(e) [see magenta and blue curves for $T$ = 8 K and 1 K, respectively]. To cancel out the background from the V-shaped DOS, we divided the intensity of symmetrized curves at 1 K by that at 8 K (see yellow curve) and we found a soft gap opening in the superconducting state with no sign of coherence peaks expected for conventional $s$-wave superconductors.

Now we discuss the origin of the unusual spectral-weight loss in the normal state. Similar behavior has been observed in several quasi-1D materials and regarded as a hallmark of TLL \cite{ARPES1, ARPES2, ARPES3, ARPES4, ARPES5}, where a confinement of correlated electrons in 1D breaks down the single-particle description \cite{TLL1, TLL2}. In the TLL picture, the DOS is significantly suppressed as $\propto$$|E|^{\alpha}$ and becomes vanishingly small near $E_{\rm F}$. The present observations of a quasi-1D band structure and a power-law-like suppression of DOS ($\alpha$ = 0.42) seem consistent with the TLL picture. However, we also found a small but finite Fermi edge which is not expected in the TLL theory. This discrepancy suggests a departure from an ideal 1D TLL state, possibly due to finite inter-chain interactions. The suppression of DOS may be also explained in terms of the formation of a pseudogap by charge-density-wave (CDW) fluctuations. While there is no clear indication for a CDW order, Tl$_2$Mo$_6$Se$_6$ would have a large CDW susceptibility because of the observed strong quasi-1D electronic structure. This is supported by the stabilization of a CDW order under uniaxial strain \cite{Strain}. Such a proximity to the ordered phase may lead to sizable CDW fluctuations and results in a pseudogap opening in DOS. In this sense, the absence of a CDW order may also be an indication of the deviation from the ideal 1D system to allow for the occurrence of robust superconductivity. Therefore, in both the TLL and CDW-fluctuation pictures, the presence of finite DOS at $E_{\rm F}$ is a consequence of the deviation from the ideal 1D system and would be responsible for the coexistence with the superconductivity. While direct evidence for a deviation from the pure-1D system may be seen as a finite band dispersion along $k_x$/$k_y$ and a wiggling of FS, possible differences in the band dispersions in Fig. 2(e) ($\sim$100 meV and $\sim$0.02 ${\rm \AA}^{-1}$) are comparable to the present experimental uncertainties, requiring higher-resolution ARPES measurements to settle this issue. Also, further complementary experiments are desired to clarify the origin of the suppressed DOS and its link to the normal-state physical properties.

Finally, we discuss implications of the present results in relation to the superconductivity in Tl$_2$Mo$_6$Se$_6$. Our ARPES measurements show that the starting point for understanding the superconductivity in the present system is a quasi-1D electronic structure with a very small DOS at $E_{\rm F}$. It is surprising that, although the BCS theory predicts that $T_{\rm c}$ decreases with the suppression of DOS (since $T_{\rm c}$ is proportional to $exp[-1/{\rm DOS}(E_{\rm F})V]$ where DOS($E_{\rm F}$) is the DOS at $E_{\rm F}$ and $V$ is the interaction potential), our sample shows a relatively high $T_{\rm c}$ value of 6.3 K. In this regard, the parameter $V$ in the BCS equation is not likely to be the cause of the high $T_{\rm c}$, because $V$ is essentially governed by the Debye energy which in Tl$_2$Mo$_6$Se$_6$ is expected to be no larger than 27 meV reported for an isostructural compound K$_2$Cr$_6$As$_6$ \cite{K2Cr6As6} consisting of lighter-mass elements; note that 27 meV is just an ordinary value for conventional low-$T_{\rm c}$ superconductors. Therefore, regardless of the origin of the small DOS is (TLL behavior or CDW fluctuations), an unconventional superconducting mechanism may be at work in Tl$_2$Mo$_6$Se$_6$.

The unconventional pairing could be playing a role in determining the shape of the DOS in the superconducting state. It has been reported that the gap anisotropy or nodes exist in various quasi-1D superconductors such as organic superconductors and metallic-chain-based systems \cite{node1, node2}. Such commonality suggests that the low dimensionality plays an important role for promoting anisotropic superconductivity. In addition, the gap anisotropy may be related to odd-parity topological superconductivity, because Tl$_2$Mo$_6$Se$_6$ is predicted to be such a topological superconductor \cite{TSC} and the origin of its topological nature (i.e., inter-orbital pairing) is similar to that in Cu$_x$Bi$_2$Se$_3$, where a clear gap anisotropy has been observed \cite{Matano, Yonezawa}. Given the quasi-1D nature and the possible odd-parity superconductivity, an anisotropic gap opening in Tl$_2$Mo$_6$Se$_6$ is not surprising, and an anisotropic gap is expected to weaken the coherence peak [note that the experimental curves in Fig. 3(e) (blue and pink circles) are not simple ARPES spectra at a specific $k_{\rm F}$ point, but are angle-integrated PES spectra to mimic the total DOS \cite{note}]. Furthermore, fluctuation effects would be enhanced at the surface in a quasi-1D superconductor (because the superconducting fluctuations in each chain are stabilized by the existence of neighboring chains), which would smear a coherence peak. Nevertheless, the lack of a coherence peak may simply be due to the lack of resolutions or unfavorable matrix elements, and one cannot make a conclusive statement at this point.

It is also remarked that the observed Dirac-cone-like band has been predicted to be responsible for the time-reversal-invariant topological superconductivity. The presence of a Dirac-cone-like band, which is protected by a nonsymmorphic screw symmetry interchanging the two sublattices A and B [Fig. 1(a)], indicates that the sublattice degrees of freedom are of crucial importance to classify superconducting states in Tl$_2$Mo$_6$Se$_6$. Theoretically, there exist six possible pairing states categorized by the difference in the sublattice symmetries, and the spin-triplet $E_{2u}$ state promoted by the inter-sublattice pairing is likely stable \cite{TSC}. Intriguingly, this topological superconducting state breaks the rotational symmetry in the $a$-$b$ plane because of a nematic order observed in Cu$_x$Bi$_2$Se$_3$ \cite{Matano, Yonezawa}. It has been also predicted that a pair of Majorana flat bands emerges on the (001) surface of Tl$_2$Mo$_6$Se$_6$. The predicted topological superconducting phase is stable over a wide parameter range even when the chemical potential is close to the present value ($\sim$0.2 eV below the calculated $E_{\rm F}$) \cite{TSC}. Therefore, Tl$_2$Mo$_6$Se$_6$ is a promising platform to realize topological superconductivity. All these intriguing aspects strongly suggest the occurrence of unconventional superconductivity in Tl$_2$Mo$_6$Se$_6$.

In conclusion, we have presented high-resolution ARPES results on Tl$_2$Mo$_6$Se$_6$ superconductor. We revealed a linear band dispersion near the BZ corner, in agreement with the presence of Dirac cones protected by nonsymmorphic screw symmetry. The observed Dirac-like energy band shows a quasi-1D characteristic such as the formation of nearly-flat FS sheets and a power-law-like suppression of the DOS upon approaching $E_{\rm F}$. We also observed a signature of an unusual DOS shape in the superconducting phase. These results lay the foundation for understanding the superconducting and topological properties of Tl$_2$Mo$_6$Se$_6$.

\begin{acknowledgements}
We thank the Helmholtz-Zentrum Berlin for the allocation of synchrotron radiation beamtime. This work was supported by Grant-in-Aid for Scientific Research on Innovative Areas ``Topological Materials Science" (JSPS KAKENHI No: JP15H05853), Grant-in-Aid for Scientific Research (JSPS KAKENHI No: JP17H04847, JP17H01139, and JP15H02105), the Program for Key Interdisciplinary Research of the Tohoku University, and by DFG (CRC1238 ``Control and Dynamics of Quantum Materials", Projects A04).\end{acknowledgements}

\newpage
\bibliographystyle{prsty}

\begin{thebibliography}{50}
\bibitem{Cuprate} P. A. Lee, N. Nagaosa, and X.-G. Wen, Rev. Mod. Phys. \textbf{78}, 17 (2006).
\bibitem{Pnictide} H. Hosono and K. Kuroki, Physica C \textbf{514}, 399 (2015).
\bibitem{Interface} Y. Saito, T. Nojima, and Y. Iwasa, Nature Rev. Mater. \textbf{2}, 16094 (2017).
\bibitem{Film} J.-F. Jia, S.-C. Li, Y.-F. Zhang, and Q.-K. Xue, J. Phys. Soc. Jpn. \textbf{76}, 082001 (2007).
\bibitem{Organic} D. Jerome, A. Mazaud, M. Ribault, and K. Bechgaard, J. Phys. Lett. \textbf{41}, L95 (1980).
\bibitem{LiMo7O11} M. Greenblatt, W. H. McCarroll, R. Neifeld, M. Croft, and J. V. Waszczak, Solid State Commun. \textbf{51}, 671 (1984).
\bibitem{A2Cr3As3} J.-K. Bao, J.-Y. Liu, C.-W. Ma, Z.-H. Meng, Z.-T. Tang, Y.-L. Sun, H.-F. Zhai, H. Jiang, H. Bai, C.-M. Feng, Z.-A. Xu, and G.-H. Cao, Phys. Rev. X \textbf{5}, 011013 (2015).
\bibitem{TMS1} J. C. Armici, M. Decroux, \O. Fischer, M. Potel, R. Chevrel, and M. Sergent, Solid State Commun. \textbf{33}, 607 (1980).
\bibitem{TMS2} R. Lepetit, P. Monceau, M. Potei, P. Gougeon, and M. Sergent, J. Low Temp. Phys. \textbf{56}, 219 (1984).
\bibitem{Anisotropy1} R. Brusetti, P. Monceau, M. Potel, P. Gougeon, and M. Sergent, Solid State Commun. \textbf{66}, 181 (1988).
\bibitem{Anisotropy2} R. Brusetti, A. Briggs, O. Laborde, M. Potel, and P. Gougeon, Phys. Rev. B \textbf{49}, 8931 (1994).
\bibitem{Calc1} A. P. Petrovi\'c, R. Lortz, G. Santi, M. Decroux, H. Monnard, \O. Fischer, L. Boeri, O. K. Andersen, J. Kortus, D. Salloum, P. Gougeon, and M. Potel, Phys. Rev. B \textbf{82}, 235128 (2010).
\bibitem{DSM} Q. D. Gibson, L. M. Schoop, L. Muechler, L. S. Xie, M. Hirschberger, N. P. Ong, R. Car, and R. J. Cava, Phys. Rev. B \textbf{91}, 205128 (2015).
\bibitem{Cubic} Q. Liu and A. Zunger, Phys. Rev. X \textbf{7}, 021019 (2017).
\bibitem{TSC} S.-M. Huang, C.-H. Hsu, S.-Y. Xu, C.-C. Lee, S.-Y. Shiau, H. Lin, and A. Bansil, Phys. Rev. B \textbf{97}, 014510 (2018).
\bibitem{Wang} Z. Wang, K. Segawa, S. Sasaki, A. A. Taskin, and Y. Ando, APL Mater. \textbf{3}, 083302 (2015).
\bibitem{SM} See Supplemental Material at [URL], which includes ref. \cite{Wang}, for details of crystal growth, experimental setup, and core-level data.
\bibitem{Calc2} T. Hughbanks and R. Hoffmann, J. Am. Chem. Soc. \textbf{105}, 1150 (1983).
\bibitem{innerpotential}The inner potential value $V_0$ is set to the energy of the VB bottom (6 eV). Note that the usual estimation of $V_0$ from a periodicity of band dispersion is difficult due to the absence of a clear band dispersion along $k_y$. Therefore, there is an ambiguity in $V_0$ and hence in the origin of $k_y$.
\bibitem{ARPES1} H. Ishii, H. Kataura, H. Shiozawa, H. Yoshioka, H. Otsubo, Y. Takayama, T. Miyahara, S. Suzuki, Y. Achiba, M. Nakatake, T. Narimura, M. Higashiguchi, K. Shimada, H. Namatame, and M. Taniguchi, Nature (London) \textbf{426}, 540 (2003).
\bibitem{ARPES2} F. Wang, J. V. Alvarez, S.-K. Mo, J. W. Allen, G.-H. Gweon, J. He, R. Jin, D. Mandrus, and H. H\"ochst, Phys. Rev. Lett. \textbf{96}, 196403 (2006).
\bibitem{ARPES3} L. Dudy, J. D Denlinger, J. W. Allen, F. Wang, J. He, D. Hitchcock, A. Sekiyama, and S. Suga, J. Phys. Condens. Matter \textbf{25}, 014007 (2013).
\bibitem{ARPES4} Y. Ohtsubo, J. I. Kishi, K. Hagiwara, P. Le F\`evre, F. Bertran, A. Taleb-Ibrahimi, H. Yamane, S. I. Ideta, M. Matsunami, K. Tanaka, and S. I. Kimura, Phys. Rev. Lett. \textbf{115}, 256404 (2015).
\bibitem{ARPES5} M. D. Watson, Y. Feng, C. W. Nicholson, C. Monney, J. M. Riley, H. Iwasawa, K. Refson, V. Sacksteder, D. T. Adroja, J. Zhao, and M. Hoesch, Phys. Rev. Lett. \textbf{118}, 097002 (2017).
\bibitem{TLL1} S. Tomonaga, Prog. Theor. Phys. \textbf{5}, 544 (1950).
\bibitem{TLL2} J. M. Luttinger, J. Math. Phys. \textbf{4}, 1154 (1963).
\bibitem{Strain} G. X. Tessema, Y. T. Tseng, M. J. Skove, E. P. Stillwell, R. Brusetti, P. Monceau, M. Potel, and P. Gougeon, Phys. Rev. B \textbf{43}, 3434 (1991).
\bibitem{node1} M. Takigawa, H. Yasuoka, and G. Saito, J. Phys. Soc. Jpn. \textbf{56}, 873 (1987).
\bibitem{node2} Z.-T. Tang, J.-K. Bao, Y. Liu, Y.-L. Sun, A. Ablimit, H.-F. Zhai, H. Jiang, C.-M. Feng, Z.-A. Xu, and G.-H. Cao, Phys. Rev. B  \textbf{91}, 020506(R) (2015).
\bibitem{K2Cr6As6} Q.-G. Mu, B. B. Ruan, B.-J. Pan, T. Liu, J. Yu, K. Zhao, G.-F. Chen, and Z.-A. Ren, Phys. Rev. B \textbf{96}, 140504(R) (2017).
\bibitem{Matano} K. Matano, M. Kriener, K. Segawa, Y. Ando, and G.-Q. Zheng, Nature Phys.  \textbf{12}, 852 (2016).
\bibitem{Yonezawa} S. Yonezawa, K. Tajiri, S. Nakata, Y. Nagai, Z. Wang, K. Segawa, Y. Ando, and Y. Maeno, Nature Phys. \textbf{13}, 123 (2017).
\bibitem{note} The experimental curves in Fig. 3(e) are obtained by integrating the ARPES intensity from $k_z$ = -0.5 to -0.9 ${\rm \AA}^{-1}$ to mimic the total DOS. We note that there must be a momentum broadening in the $k_y$ direction (perpendicular to the sample surface), which is inevitable in the ARPES measurements with vacuum-ultraviolet photons as in the present case. Considering these facts, the coherence peak may be substantially broadened in the curves as those in Fig. 3(e) when the gap size is highly anisotropic in the momentum space.

\end{thebibliography}

\end{document}